\documentclass[12pt,preprint]{aastex}
\usepackage{amssymb, amsmath}

\shorttitle{Binary Frequency of Wolf-Rayets in M31 and M33}

\shortauthors{Neugent et al.}

\begin{document}

\title{The Close Binary Frequency of Wolf-Rayet Stars \\as a Function of Metallicity in M31 and M33\altaffilmark{1}}

\author{Kathryn F. Neugent and Philip Massey}
\affil{Lowell Observatory, 1400 W Mars Hill Road, Flagstaff, AZ 86001; \\kneugent@lowell.edu; phil.massey@lowell.edu}

\altaffiltext{1}{The spectroscopic observations reported here were obtained at the MMT Observatory, a joint facility of the University of Arizona and the Smithsonian Institution. MMT telescope time was granted by NOAO, through the Telescope System Instrumentation Program (TSIP). TSIP is funded by the National Science Foundation. This paper uses data products produced by the OIR Telescope Data Center, supported by the Smithsonian Astrophysical Observatory.}

\begin{abstract}
Massive star evolutionary models generally predict the correct ratio of WC-type and WN-type Wolf-Rayet stars at low metallicities, but underestimate the ratio at higher (solar and above) metallicities. One possible explanation for this failure is perhaps single-star models are not sufficient and Roche-lobe overflow in close binaries is necessary to produce the ``extra" WC stars at higher metallicities. However, this would require the frequency of close massive binaries to be metallicity dependent. Here we test this hypothesis by searching for close Wolf-Rayet binaries in the high metallicity environments of M31 and the center of M33 as well as in the lower metallicity environments of the middle and outer regions of M33. After identifying $\sim$100 Wolf-Rayet binaries based on radial velocity variations, we conclude that the close binary frequency of Wolf-Rayets is not metallicity dependent and thus other factors must be responsible for the overabundance of WC stars at high metallicities. However, our initial identifications and observations of these close binaries have already been put to good use as we are currently observing additional epochs for eventual orbit and mass determinations.
\end{abstract}


\section{Introduction}
Wolf-Rayet (WR) stars are readily identifiable by their characteristically broad emission lines. These lines form in the star's accelerating outer layers, where strong stellar winds have pushed away the hydrogen-rich outer layers of an O-type star to create a WR star. The type of WR star (WN - nitrogen rich, or WC - carbon rich) then depends upon which layer is visible based upon how much mass has been lost. WN-type WRs have lost enough mass for the H-burning products, nitrogen and helium, to dominate the spectrum. Further mass-loss results in a WC-type WR, where the He-burning products carbon and oxygen dominate. The impetus for this mass-loss depends upon whether the star is single or a member of a close binary system. 

For single stars, a WR forms from an O-type star through the ``Conti scenario" as a result of its strong stellar winds (Conti 1975, Maeder \& Conti 1994). These stellar winds are driven by radiation pressure on highly ionized metal lines, and hence the mass-loss rates are metallicity dependent. Early comparisons of the WR content of the Magellanic Clouds and the Milky Way revealed a strong metallicity dependence in the relative number of WCs and WNs, with the metal-poor SMC being dominated by WNs, while the solar neighborhood had roughly an equal number of these stars. Furthermore, proportionately more WCs were found towards the Galactic center, where the metallicity is higher (Smith 1968). Vanbeveren \& Conti (1980) argued that these differences were due to the effect of metallicity on the mass-loss rates, causing WCs to form earlier in metal-rich systems, and hence be more numerous. (Smith 1973 had earlier argued that metallicity might be responsible for the relative absence of WCs in the Magellanic Clouds, but without understanding the physical mechanism.) Eventually a galactrocentric gradient in the relative number of WCs and WNs was also discovered in M33 (Massey \& Conti 1983) further adding suspicion that the relationship between metallicity and stellar winds was responsible for changes in the evolved massive star populations seen from system to system, although other possible explanations, such as changes in the initial mass function, could not be dismissed.

The same stellar-wind mechanism might be responsible for O stars becoming WRs even in binary systems (Conti 1975), and indeed Massey et al.\ (1981) argued that most WRs form as a result of mass-loss from stellar winds based upon the relative frequency of unevolved and evolved systems. Still, {\it some} WR stars must have formed as the result of close binary evolution with Roche-lobe overflow (RLOF) playing a dominant role in the mass loss. The evolution of such systems was well described by van den Heuvel (1973, 1976), de Loore \& De Gr\`{e}ve (1975), de Loore et al.\ (1975), and Vanbeveren \& Conti (1980), amongst others. Many (but not all) massive close binaries have components with similar masses, and hence should evolve somewhat in tandem (Garmany et al. 1980, Kobulnicky \& Fryer 2007).  The (slightly) more massive star evolves to a blue supergiant, expanding in radius. If the star reaches its Roche surface, enhanced mass-loss quickly causes it to become a WN-type Wolf-Rayet star. If the initial masses were sufficiently similar, the luminosity of the secondary will be comparable to that of the WR in the visible, and the absorption lines should be readily seen, {\it if} the spectral resolution and signal-to-noise is good enough. (Vanbeveren \& Conti 1980 argue this should usually be the case based on the Garmany et al.\ 1980 results.) Such a WR binary is thus ``double-lined" and spectroscopically it will be a WN+OB star. Of course, the orbital motion will result in the emission lines (from the WR star) and the absorption lines (from the OB star) moving in anti-phase. If the luminosities aren't compatible, or the signal-to-noise is too poor (as we expect for most of the faint WR stars we observe here) then the system will be single-lined, with only the emission of the WR visible. Further evolution will lead to a WC+OB system. The WC star will eventually explode as a supernova, leaving behind either a neutron star or a black hole. If too much mass has been lost as a result of the explosion, the system will no longer be bound, and the (former) OB secondary will become a runaway star as a result. But much more likely, the system will remain bound and the OB star plus a compact companion will become an x-ray source since the OB star's stellar wind impinges upon the compact object.  Evolution of the OB star will next lead to a WN stage, again aided by RLOF if the OB star expands sufficiently. This WN star will be a ``single-lined" binary and subsequent evolution will lead to a single-lined WC binary. There are of course several possible variants on this scenario: one can imagine that if the initial masses are nearly identical that a WR+WR binary will be produced before the more massive star undergoes a SNe explosion. Indeed there were several WRs classified as ``WN+WC" in the catalog of Galactic WRs by van der Hucht et al.\ (1981) stars, but Massey \& Grove (1989) demonstrated that the C and N lines move in phase in two such stars. No WR+WR systems have yet been identified (cf.\ van der Hucht 2001).

Meynet \& Maeder (2005) found that the Geneva evolutionary models are able to correctly predict the ratio of WC-type and WN-type WRs at low metallicities, but underestimated the ratio at metallicities of solar and higher. The Geneva evolutionary models do not (yet) include the effects of binary evolution, but Vanbeveren et al.\ (2007) and Eldridge et al.\ (2008) found that evolutionary models that contain some fraction of binaries can better reproduce the WC to WN ratio at higher metallicities. However, since that time, there have been substantial improvements in the ``observed" ratios thanks to our deeper and more complete surveys (Neugent \& Massey 2011, Neugent et al.\ 2012). Currently the data suggest that the single-star Geneva models do a good job with the lower metallicity cases (such as in the LMC and outer regions of M33), but predict too few WC stars relative to the number of WNs at metallicities of solar and above (Neugent et al.\ 2012). Neugent et al.\ (2012) note that one possible explanation is that the binary evolution (i.e., mass-loss via RLOF) might play a more important role in forming WRs at higher metallicities than at lower (see also Georgy et al.\ 2012). If this hypothesis is correct, it would suggest that the close binary frequency of WRs is metallicity dependent. Here we aim to test this by identifying close WR binaries within the high and low metallicity regions of M31 and M33 and determining whether the close binary frequency varies with metallicity. Hints that this metallicity dependence may exist have been discussed before (see Zinnecker 2003) and sufficiently little is known about massive star binary formation that this dependency can't be ruled out.

While we cannot determine an absolute frequency of close WR binaries in M31 and M33 based upon a single year of observations, we can instead determine if the {\it relative} frequency of close binaries depends upon metallicity. M33 has a strong metallicity gradient, with $\log(\rm O/H) + 12 = 8.7$ at the center and $\log(\rm O/H) + 12 = 8.3$ in the outer regions (see Magrini et al.\ 2007 and discussion in Neugent \& Massey 2011). The metallicity in M31's star-forming disk is relatively high, with $\log(\rm O/H) + 12 = 8.9$ (Zartisky et al.\ 1994, Sanders et al.\ 2012). Thus, comparing the relative binary frequency in M31 with that in the inner and outer regions of M33 allows us to show whether or not the Geneva evolutionary model's tendency to underestimate the relative number of WC to WN stars at high metallicity is due to their lack of inclusion of binaries or whether the problem exists elsewhere. 

In Section 2 we describe our observing campaign, in Section 3 we discuss newly discovered M33 WRs and in Section 4 we explain how we used radial velocities to identify the close WR binaries. In Section 5 we discuss our results and finally, in Section 6 we summarize our findings and future goals.

\section{Observations and Reductions}
Recent surveys by Neugent \& Massey (2011) in M33 and Neugent et al.\ (2012) in M31 have brought the number of known WRs within these two galaxies to 206 and 154, respectively. Both of these surveys sampled the entire galaxy and were complete to $\sim$5\%. We were thus able to use these unbiased samples to craft our candidate list. In the end, we observed 250 stars: 106 in M31 and 144 in M33. 

Our ability to undertake this project was in a large part due to the existence of the multi-object fiber-fed spectrograph Hectospec (Fabricant et al.\ 2005) on the 6.5-m MMT. Its large field of view ($1^{\circ}$ in diameter) was well matched to our survey areas of M31 and M33. Hectospec's 300 fibers and their allowed close spacing ($20\arcsec$) let us observe a multitude of candidates using only 4 pointing configurations. Finally, Hectospec's queue observing mode allowed us to request observations of the same configurations at multiple times throughout the semester to enable our search for radial velocity variations. 

We were assigned 2.5 nights of dark time in the Fall of 2012 through NOAO (2012B-0129). When designing the fiber configurations, we were able to assign 71\% of our M31 WRs using two configurations and 77\% of our M33 WRs using an additional two configurations, making a total of four configurations. Two pointings were needed in M31 for areal coverage while two more were needed in M33 because of crowding. These configurations were then observed four times (except M33 \#2, which was only observed 3 times) on eleven different nights throughout the semester. Table~\ref{tab:Obs} shows the dates each configuration was observed. Observations on 2012 Nov 8 were taken under poor observing conditions and a few of the spectra were thus unusable. The data were taken with the 270 line mm$^{-1}$ grating, resulting in spectral coverage from 3700-9000~\AA. The 250$\mu$m fibers (each subtending 1\farcs5 on the sky) resulted in a spectral resolution of 6\AA\ (5 pixels). While we expect the spectra to be contaminated by second-order blue light beyond $\sim$7500~\AA, we did not use any lines in this contaminated region. Reductions were then carried out through the standard Hectospec pipeline (Mink et al.\ 2007) by Susan Tokarz of the OIR Telescope Data Center.  The typical (median) signal-to-noise was 25 per spectral resolution element in the continuum, but varied from 10 to $>100$ depending upon the star and observing conditions. Further details about the calibration, flat-fielding and reduction procedures can be found in Drout et al.\ (2009).

\section{Newly Discovered M33 Wolf-Rayets}
In our recent M33 and M31 surveys (Neugent \& Massey 2011, Neugent et al.\ 2012) there were 11 WR candidates that we did not have a chance to confirm spectroscopically: 6 in M33 and 5 in M31. We tried to take advantage of spare fibers to remedy this. We were able to observe 5 of the remaining M33 candidates as part of the present study, and recently obtained spectra of the sixth star as part of our follow-up study of the binaries we identify here. Unfortunately none of the M31 candidates were assigned either due to crowding or location outside of the Hectospec fields. We list the results in Table~\ref{tab:candidates}. We find that five of the six M33 stars are WRs; the other is an Of star. We include in Table~\ref{tab:candidates} the equivalent widths (EWs) and full width at half maximum (FWHM) of He II $\lambda 4686$ and CIV $\lambda\lambda 5802-12$ as well as the de-projected radial distance within the plane of M33 ($\rho$). Additionally, we include the absolute magnitudes and membership in OB associations, consistent with the information we made available for the other M33 WRs in Tables 1 and 5 of Neugent \& Massey (2011).

\section{Identifying the Binaries}
With such a scant number of observations (typically 4-6), and possible unfavorable inclinations, we can't hope to identify all of the WR binaries in our sample. Still, we can see if the fraction of WR stars with significant radial velocity variations changes based on location, and thus metallicity.

There are many challenges in obtaining radial velocities for WR stars.  First and foremost, the lines are extremely broad, typically a few {\it thousand} km~s$^{-1}$ in width.  Massey (1980) found that he could achieve radial velocity precision of about 20 km~s$^{-1}$ using (high-dispersion) photographic spectra as long as an objective criteria (such as a centroid) was employed. Here we have considerably higher signal-to-noise data obtained with a linear device, and (as we show below) manage to achieve measuring errors of the order of 1-5 km~s$^{-1}$.

Yet another challenge when measuring a WR's strong emission lines deals with line profile variability in WR stars as discovered by Moffat et al.\ (1988). These variabilities are caused by narrow emission bumps superposed on the broad emission profile, giving evidence that the stellar winds are not homogenous. Instead, they are ``clumped" as Fullerton et al.\ (2006) later showed. Because of the ``clumped" winds, profiles of emission lines are constantly changing, making them difficult to accurately measure.

The third challenge is that we cannot average the radial velocities of multiple emission lines measured from the same spectra. This is due to how the emission lines are formed within the WR star's expanding atmosphere (see, for example, Hillier 1991). The stellar winds that create the WR's emission lines are accelerating and since the emission lines are formed at different places within the stellar winds, their velocities vary based on where they were created. Additionally, many of the emission lines are blends and their effective rest wavelengths haven't been well determined (Beals 1930, Kuhi \& Sahade 1968). Additionally, there can also be radiative transfer effects such as electron scattering (particularly for He II $\lambda 4686$) that further complicate the situation (Auer \& van Blerkom 1972, Hillier 1984).  Thus, we must treat the radial velocity of each line separately.

Finally, once we do identify a subset of binaries, we must consider their colliding winds. As discussed in Flores, et al. (2001), these collisions produce profile distortions within emission lines that vary with orbital phase. This has been shown to cause up to a $\sim$12\% change in the equivalent width of prominent emissions lines in V444 Cyg, a WR + O binary system (Flores et al.\ 2001). Additionally, as shown by Foellmi et al.\ (2008) for HD 5980 (a WR/LBV binary), these wind-wind collisions increase centroid-measurement error since the emission lines become asymmetric as the orbital cycle changes.

\subsection{Radial Velocities}
Our next task was to determine which of our 250 observed candidates exhibited large enough scatter in their radial velocities to indicate their presence in a close binary system. Before measuring the radial velocities, we normalized the spectra using a 4th order cubic spline after defining continuum regions by hand. We then restricted our measurements to the strongest lines in each WR, i.e., N III $\lambda\lambda 4630-34-41$, C III $\lambda 4650$, He II $\lambda 4686$, He II $\lambda 5411$, and CIV $\lambda\lambda 5802-12$, depending upon which were present. In addition to measuring all of the spectra observed as part of this observing project, we also measured the ``discovery" spectra we originally used to classify these stars as described in Neugent \& Massey (2011) and Neugent et al.\ (2012).

We initially measured the centroid of all visible emission lines in each spectrum using ``splot" in IRAF\footnote{IRAF is distributed by the National Optical Astronomy Observatory, which is operated by the Association of Universities for Research in Astronomy (AURA) under cooperative agreement with the National Science Foundation.} and defining the continuum on either side of the line by eye. To get a sense of our internal error, we measured 100 lines 3 or 4 times each. After doing these measurements, it became clear that our internal error depended on the line flux. We were thus able to determine a relationship between the errors and the line flux. We found that the emission lines with high line fluxes had internal errors of 1 km s$^{-1}$ or less and emission lines with low line fluxes (which were not used in the final calculations, as discussed later) had internal errors of around 5 km s$^{-1}$. We were then able to use these wavelength measurements to determine the radial velocities of individual lines in each spectrum.

As an independent verification of our hand-calculated radial velocities, we additionally calculated the radial velocities of each spectrum using the IRAF cross-correlation package ``fxcor." To do this, we cross-correlated all of the spectra for a particular candidate against itself. This allowed us to then use the error provided by ``fxcor" as the internal error. On average, we found our internal errors to be quite high at around 5 km s$^{-1}$. However, this makes sense given the normal usage of ``fxcor." Cross-correlation techniques are highly effective when there is a high density of spectral lines. But in the early-type stars that we are looking at, the lines are sparse and continuum dominates. Therefore, we chose to do the cross-correlation on a line by line basis. Additionally, line profile variability increases the internal error when using cross-correlation techniques because the profile shape will differ slightly between the template and observation as the bumps move around. So, while our velocities given by ``fxcor" are valid, in some ways our hand measurements might be more accurate. Either way, as we discuss later, our hand measurements and the velocities given by ``fxcor" agree nicely.

A complete list of radial velocity measurements will be published in a later paper after we have obtained enough information to compute orbits and masses of the binary systems (and thus, have obtained even more radial velocity measurements). 

\subsection{Calculating $E/I$}
To determine which of our candidates are close binaries, we needed to look for those with statistically significant radial velocity variations. This is often done by comparing the internal errors ($I$) with the external scatter ($E$) since stars with large external radial velocity scatter relative to a small internal error are likely to be close binaries. Usually $I$ is estimated from how well the radial velocities of the various lines agree in the same spectrum, but as we explained above, this method doesn't work for WR stars as we expect different lines to be formed in different layers of the star. Thus, these lines will have different intrinsic velocities and can't be used to compute $I$. So, instead we use our estimate of the internal error as described in the section above. 

When calculating our final $E/I$ value for each star, we used the radial velocity measurements of the strongest emission line in the spectrum. These lines are listed in Table~\ref{tab:Sigma}. However, other strong emission lines gave similar values for $E/I$. For the actual calculation of $E/I$ from the radial velocities $v$ for each star, we used the standard deviation $\sigma$ of $v$ for $E$ and our measurement error estimate averaged for each star as $I$. Thus, our equation is as follows: 
\begin{equation*}
E/I = \frac{\sigma(v)}{{\rm avg}(v_{\rm{err}})}.
\end{equation*}

We also computed $E/I$ using the cross-correlation (``fxcor") measurements.  For each star we produced multiple $E/I$ values, as we used each spectrum in turn as the template, cross-correlating against all of the others. We then computed a mean $E/I$, weighing the individual values inversely by the square of the average fitting error. Finally, note that in our case, our values for the internal error were all comparable (between $1-5$ km s$^{-1}$), meaning it would be unlikely for us to miss a binary with high external scatter $E$ due to our own poor measuring error $I$.

We were then able to compare the $E/I$ results both from our hand and cross-correlation measurements of the emission lines, as shown in Table~\ref{tab:Sigma}. Also shown in Table~\ref{tab:Sigma} is the identity of the emission line we used, as well as the number of spectra $N$.

As emphasized in the seminal study by Garmany et al.\ (1980), the $E/I$ test we used is a simplified version of a more general analysis of variance (AOV) test.  In the AOV test one can consider if differences in radial velocities from one observation to another are significant using multiple lines, even if there are systematic differences in the radial velocities of individual lines as we would expect here. (The same situation can occur with O stars, as even the absorption lines can have differing radial velocities due to these stars' atmospheric extent and the acceleration of stellar winds even down in the photosphere.) However, instead we choose to concentrate our analysis on a single strong line given our very modest signal-to-noise ratio and our realization of how quickly the internal errors in our measurements grew as the line flux decreased, as discussed above.

To determine whether there were systematic errors caused by our multiple observing runs, we examined the radial velocities of $\sim$50 ``single" stars ($E/I < 2$) and averaged the differences between radial velocity results for all four configurations. These differences were $\sim$1 km s$^{-1}$ and thus we conclude that there were no significant night to night or configuration to configuration variations.

\begin{figure}
\epsscale{0.5}
\plotone{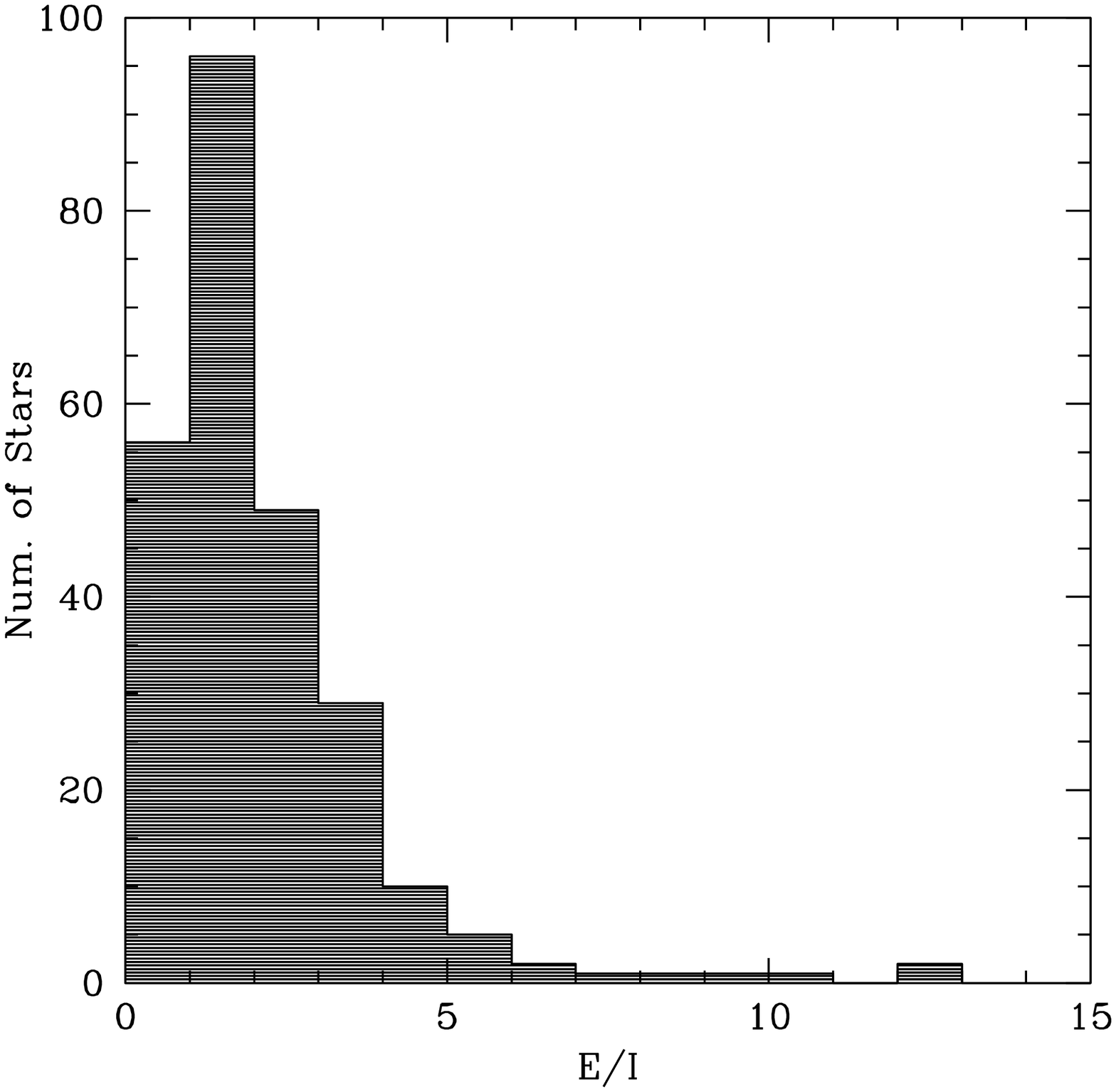}
\caption{\label{fig:sigmaHist} Histogram of $E/I$ values. In the ideal case, we expect non-binaries to have $E/I$ values of order 1. However, in practice, we adopt a $E/I$ value of $> 2$ as the dividing line between close binaries and non-binaries, as we describe in the text.}
\end{figure}

We additionally investigated whether our internal errors of 1 km s$^{-1}$ for our hand-measurements and 5 km s$^{-1}$ for our cross-correlation measurements are realistic. While a small amount of underestimation is certain possible (see Caldwell et al.\ 2009), we believe a large amount is unlikely for three reasons. First, as described above, we calculated $E/I$ using two drastically different methods. When using cross-correlation, we relied on the formal errors as our internal errors. Then, when hand-measuring the radial velocity shifts, we instead developed a relationship between the error and the line flux. After using these two methods, our results were quite self-consistent. This would not be the case if our internal errors were vastly underestimated in one case and not the other. Secondly, it might be possible for our internal errors to be underestimated in both cases. However, we don't believe this has occurred. If our internal errors were off by a factor of 5 (for example), then the peak of the $E/I$ histogram shown in Figure~\ref{fig:sigmaHist} (and discussed further in the next section) wouldn't be so close to one. Instead, the peak would fall at the unrealistic value of $E/I << 1$. Finally, we observed $\sim$10 non-binary stars ($E/I < 2$) twice in the same night (with a gap of 3 hours between observations). We then compared our velocity calculations for these stars and found that the average difference in radial velocity was $0.95\pm1.1$ km s$^{-1}$. Therefore, we believe that our internal errors are accurately represented.

\section{Results}
\subsection{Binary Frequency as a Function of Metallicity}
In the perfect world, our distribution of $E/I$ values would be bimodal with all of the single-star systems having $E/I$ values of $\sim$1 and all of the close binary systems having $E/I$ values that are much larger. But, as Figure~\ref{fig:sigmaHist} shows, this bimodal distribution does not exist. Instead, we see a unimodal distribution centered around $E/I = 1$ with a tail extending towards $E/I > 1$. In practice, one usually adopts a $E/I$ value of $> 2$ as the dividing line between close binaries and non-binaries (Abt \& Levy 1976, Abt 1987) since internal errors may be underestimated.

Table~\ref{tab:BinFreq} shows the frequency of binaries that we detected among the M31 and M33 WRs, where we've broken M33 up into three separate regions based on its varying metallicity (see Neugent \& Massey 2011 for a full explanation). As shown in Table 4, the percentages of close binaries agree across all regions to within a few percent, with the exception of the inner portion of M33. In order to test the robustness of this conclusion, we also include the percentages based upon cutoff values for $E/I = 3$ and 4. While the percentages go down, it's obvious that the relative values stay the same\footnote{To understand how much the percentages are affected by small number statistics, we provide uncertainties based upon Poisson statistics, i.e., that the variation in either the number of binaries or the total number of WRs might vary by the square root of the quantity.  Thus, these uncertainties should not be over-interpreted but are provided purely as a means of evaluating to what extent small-number statistics might affect the results. We used this method to good effect in evaluating the WC/WN ratios as a function of metallicity in nearby galaxies; e.g., Neugent \& Massey (2011).}. 

Given that the metallicity of the inner portion of M33 is in between that of M31 and the middle region of M33, it is surprising that its binary frequency does not agree better. One possible explanation is that the inner region of M33 is spatially condensed compared to other M33 regions we consider.  One of the assumptions in all studies such as ours is that the star-formation rate is has not changed significantly over the relevant evolutionary time scales. We have emphasized in our previous papers that this should be a good approximation when discussing galaxy-wide stellar populations, but will break down on small spatial scales where a recent burst of star formation could affect the results. Regardless, we find that the fraction of close binaries is {\it lower} in the inner region of M33, so this is in the opposite sense of what would be the case if binarity was responsible for the overabundance of WCs.

A more exacting test is provided by asking what fraction of WCs are close binaries in these four regions. Again, we must recall that just because a WC is not detected as a binary today doesn't mean that it did not form through RLOF: we may fail to detect such a system either because the inclination was unfavorable or (for that matter) that its initial companion lost sufficient mass to eject the WC as a runaway.  But, our comparison is intended to be differential: given that the models do well with predicting the WC/WN ratios at low and intermediate metallicities, do we find a higher fraction of WCs as (present-day) binaries in our sample at high metallicities?  We provide the fraction of WNs and WCs that we detected as binaries (using $E/I>2$) in Table~\ref{tab:BinFreq} and again we see the answer is no.  The data are consistent with the close binary frequency of WC stars being identical in all four regions, and, if anything, shows a trend in the opposite sense. We also find that the deficiency of WR binaries in the central part of M33 is not due to fewer WCs having companions, but rather a lower fraction of WN binaries being found. 

If the ``extra" WCs (relative to the number of WNs) found in high metallicity environments is not due to binarity, what then is the explanation? The models at high metallicity would have to increase the duration of the WC stage or decrease the duration of the WN stage. Higher mass-loss rates during the Luminous Blue Variable (LBV) phase would possibly allow this, or decreased mixing. Increasing the mass-loss rates during the red supergiant phase might also affect the WC/WN ratio favorably, although this is less clear.  But again one is faced with the question of why these would be metallicity dependent.  (We are grateful to Georges Meynet and Cyril Georgy for comments on this point.) 

We will note that the problem at high metallicity could be alleviated, and perhaps eliminated, once revised Geneva models become available for extra-solar metallicities. The latest generation of evolutionary models from the Geneva group include many improvements (see, e.g., Ekstr\"{o}m et al.\ 2012, Georgy et al.\ 2012, 2013), but are so far only available at metallicities up to solar.  Efforts are underway to compute grids of models at the $2\times$ solar metallicity we expect in M31 (C. Georgy 2014, private communication), and it will be intriguing to compare the WC/WN ratios from those models with our data.  The central part of M33 (where the metallicity is approximately solar) does show a higher WC/WN ratio than predicted by the models, but as we argue above, that region is the least robust to the graininess of the star formation rate.  

The purpose of our study was to examine the relative binary frequency in M31 and M33 and not to determine the absolute binary frequency. To compute the absolute binary frequency, we would need not only many more observations for better phase coverage, but then also to correct the frequency for inclination effects. In terms of understanding the role binary evolution has played, we would also have to account for the cases where the WR star formed as part of a binary but became a runaway star when its companion exploded as a supernova. Still it is interesting to compare our results with what is known elsewhere. For instance, Foellmi et al.\ (2003) argues that the ``true" (corrected) percentage of close WR binaries among the early-type WNs in the Magellanic Clouds is $30-40$\%. This is comparable to what is known in the Milky Way (see discussion in Massey 1981). As Table~\ref{tab:BinFreq} shows, our results agree relatively well.

\subsection{Going Further: A Demonstration Project}
\begin{figure}
\epsscale{1}
\plotone{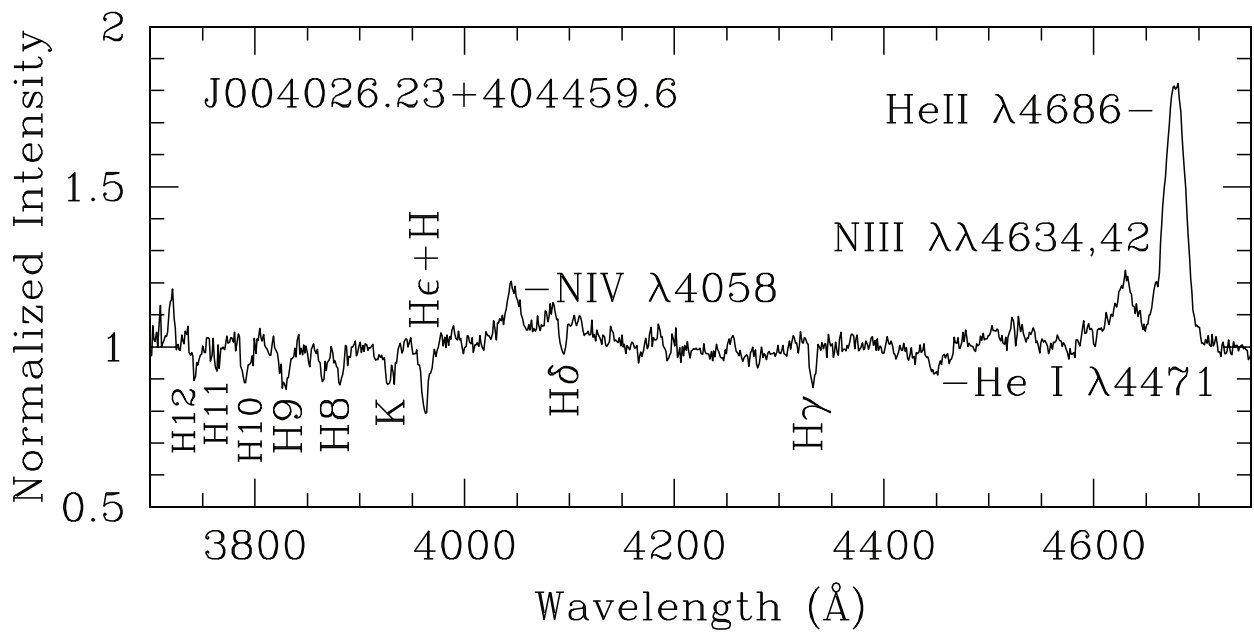}
\caption{\label{fig:emiAbs} Spectrum of M31 star J004026.23+404459.6. The emission comes from the WR star (WN5), the absorption from an OB companion (roughly B0). Interstellar Ca II H and K lines are also evident.}
\end{figure}

\begin{figure}
\epsscale{0.49}
\plotone{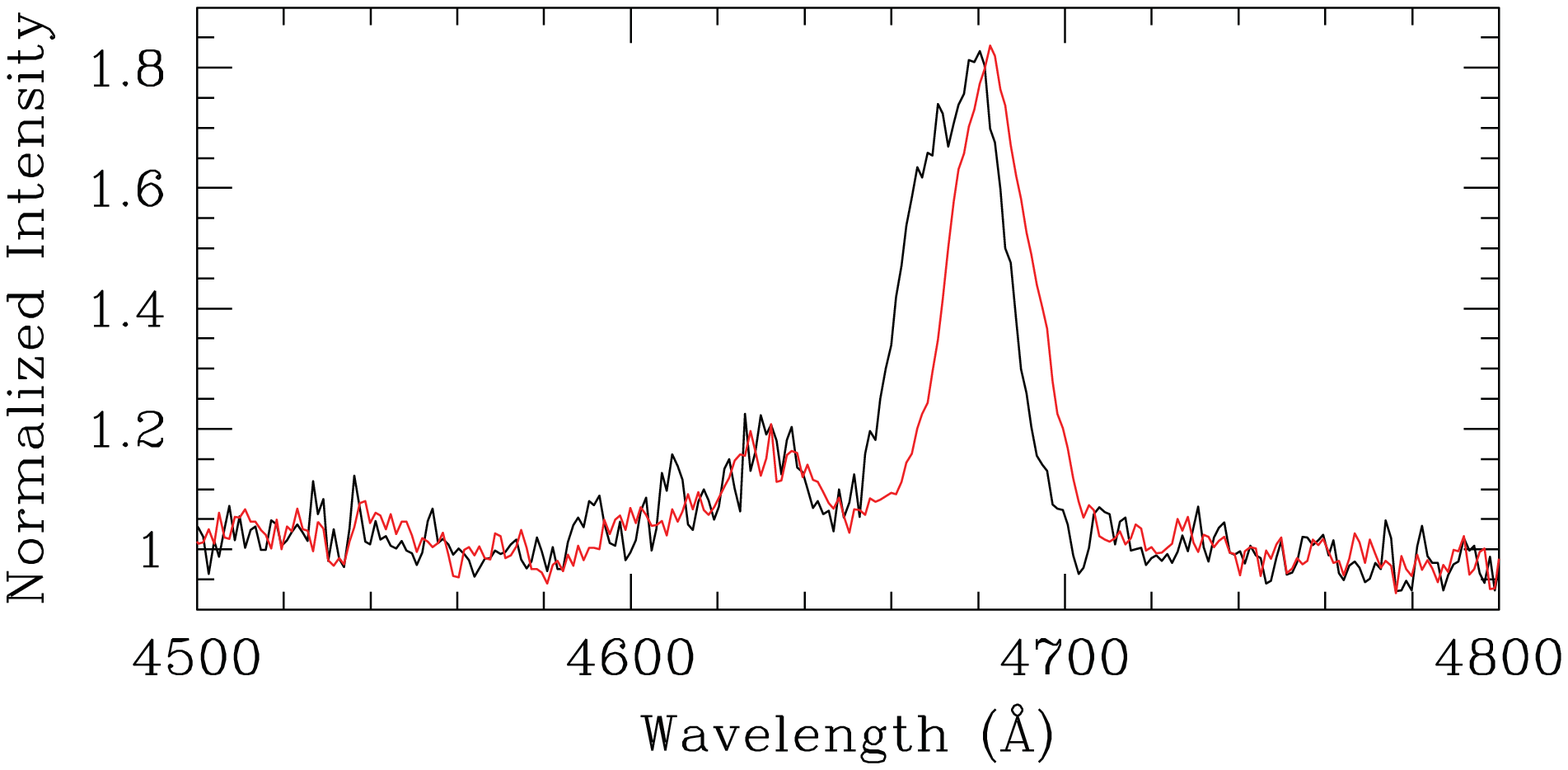}
\plotone{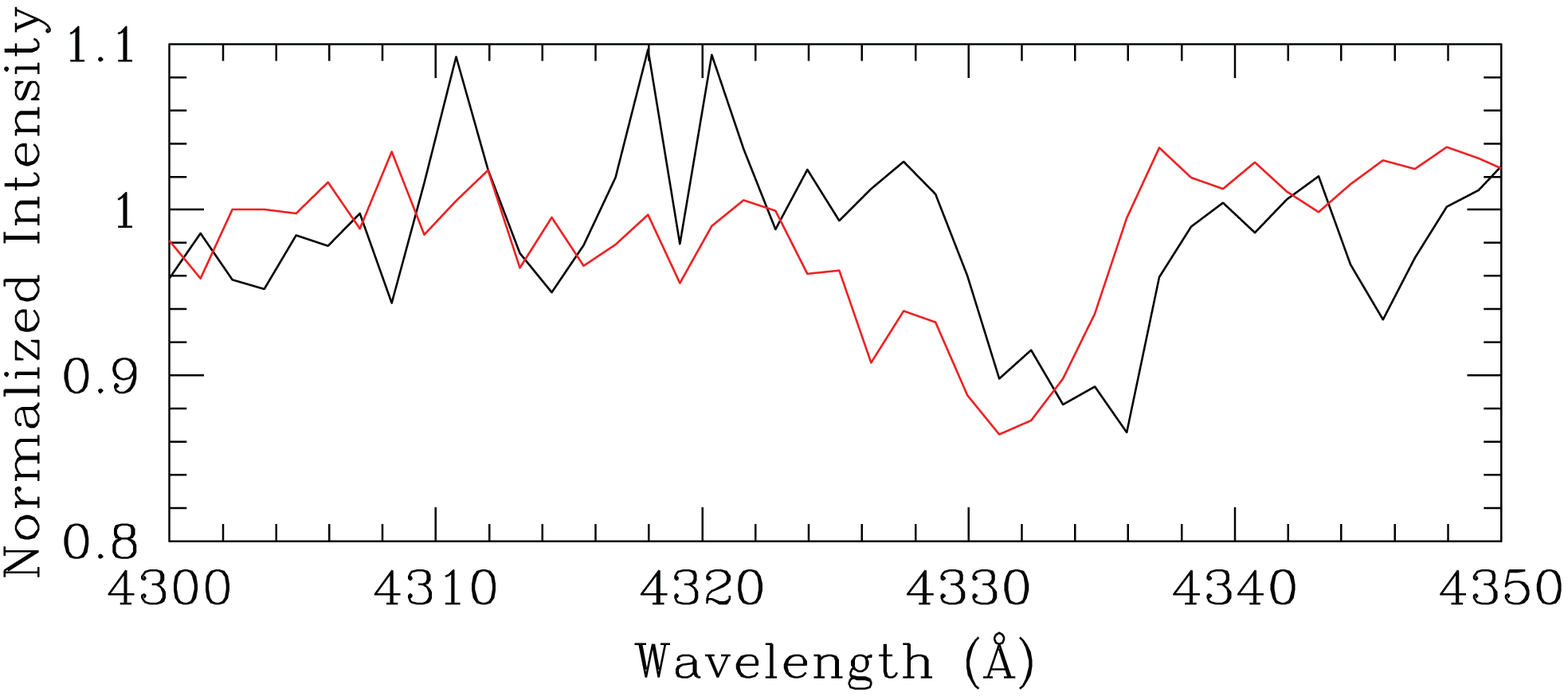}
\caption{\label{fig:spec} Radial velocity of emission (left) vs.\ absorption (right) in J004026.23+404459.6. Black corresponds to our MMT spectrum obtained on 2012 Nov 8; red to the one on 2012 Dec 11.}
\end{figure}

In the process of studying the relationship between binary frequency and metallicity, we've also identified many potential new close WR binaries: 102 stars with $E/I > 2$, 56 with $E/I > 3$, and 23 with $E/I > 4$. One of our new found binaries is J004026.23+404459.6, a WN+OB star in M31 with an $E/I$ value of 4.44 and strong upper Balmer lines. Its spectrum is illustrated in Figure~\ref{fig:spec}. Strong He II $\lambda$4686, N III $\lambda\lambda$4630-34-41 and N IV $\lambda$4058 come from the WR star, while the He I $\lambda$4471 absorption and the Balmer absorption lines (H$\gamma$, H$\delta$, H$\epsilon$, and H8-12) all come from an OB companion. We see in Figure~\ref{fig:emiAbs} that the emission and absorption components both change in radial velocity in the opposite sense; these two spectra were taken around a month apart and show the maximum velocity separation we see from our 6 spectra. Of course, with only 6 observations, we don't know the period, and we certainly cannot obtain an orbit or masses. But, even so, we can learn something interesting about the formation of WRs from these data. 

\begin{figure}
\epsscale{0.5}
\plotone{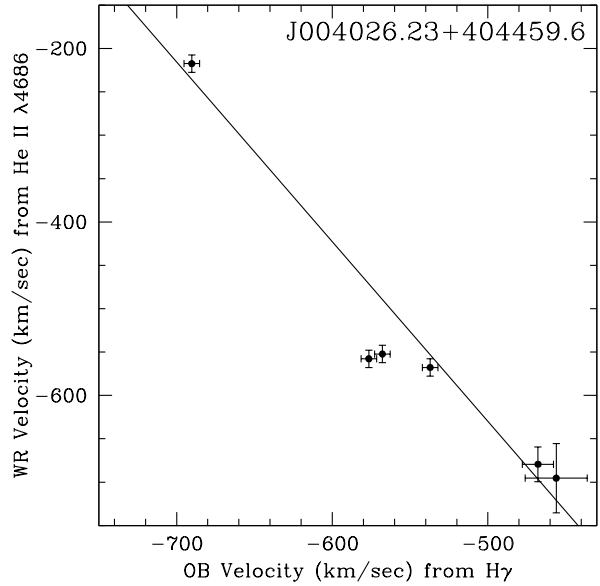}
\caption{\label{fig:Wilson} Wilson Diagram for J004026.23+404459.6. The radial velocity of the He II $\lambda$4686 emission line (from the WR star) is plotted against the radial velocity of the H$\gamma$ absorption line (from the OB star). The least-squares linear fit is shown. Its slope, $2.27 \pm 0.07$, means that the WR star has a mass that is only 44\% of its OB companion. Yet, it must have started its life as the initially more massive star.}
\end{figure}

Wilson (1941) noted that if you plot the radial velocity of one binary component against the radial velocity of the other component, the points should fall on a straight line, the slope of which is the inverse of the mass ratio of the two stars. We show such a ``Wilson diagram" in Figure~\ref{fig:Wilson}. The slope is $2.27 \pm 0.07$, implying that the WR star has a mass that only about 44\% as much as that of the OB star. Yet, from a stellar evolutionary point of view, we expect that the WR star must have begun as the more massive component, suggesting that it must have lost more than half of its mass in becoming a WR. An orbit solution would tell us the orbital separation and whether the stars are filling their Roche lobes, and whether this is possibly just a case of binary evolution, where the OB star has excreted mass from the progenitor of the WR star as the system evolved. But, were we to find that these stars are relatively well separated, we would know that the WR progenitor has lost this mass by other means. Is this mass loss consistent with stellar winds? To answer this, we need the mass, as the expected mass-loss rates depend upon the mass (luminosity). If the system is massive and luminous enough, it's possible. But, if it's not, it would then require mass loss during the LBV phase to explain its evolution. However, as we describe below, we've begun an observing campaign to directly answer these questions for this, and many of the other systems we've discovered.

\section{Conclusions and Future Work}
Out of our 250 WRs in M31 and M33, we found 102 stars with $E/I > 2$, 56 with $E/I > 3$, and 23 with $E/I > 4$. These close binaries are found throughout M33 and M31, with a binary frequency independent of location, except in the center of M33 where fewer WN binaries were found. The fraction of WC close binaries appears to be the same in all high and low metallicity regions we've examined. Thus, we can conclude that a larger binary frequency at high metallicities is not responsible for the discrepancy with the models. Somehow the models need to predict longer life spans for WCs and/or lower life spans for WNs at high metallicities.

As discussed in the previous section, our path ahead is clear. Armed with around 100 newly discovered WR and O-star binaries, we plan on determining the systems' orbits and masses. Recent work suggests that the stellar winds driving a massive star's mass loss are not as powerful as previously thought (see Puls et al.\ 2008), and that other mechanisms (such as episodic mass loss during the LBV phase or RLOF in close binary systems) may be responsible for stripping down the star (Smith \& Owocki 2006, Smith 2012). By directly measuring the orbits and masses, we will be able to determine whether the ``normal" mass-loss rates from stellar winds are sufficient to produce what we observe or whether other explanations are needed. At this point we've just finished getting the second epoch (this paper described the first) of radial velocity observations using Hectospec on the MMT. However, we still need another two seasons of observations for orbit solutions. Additionally, we plan to begin monitoring these stars photometrically to obtain orbital inclinations (needed for the masses) using Lowell Observatory's new 4.3-m Discovery Channel Telescope. Hopefully with this new information we will be able to provide unprecedented observational constraints on the evolutionary models of massive stars, and help answer how Wolf-Rayet stars form.

\acknowledgements
We would like to thank Grant Williams for his help in scheduling the observations suitably, Perry Berlind and Mike Calkins for their assistance while observing, Susan Tokarz for reducing the spectra, Nelson Caldwell for his help in observing and scheduling our configurations, and Georges Meynet, Cyril Georgy, and an anonymous referee for helpful comments that improved the paper. This work was supported by the National Science Foundation under AST-1008020.

\clearpage

\begin{deluxetable}{l c c c c c c c c c}
\tablecaption{\label{tab:Obs} Dates (UT) of New Observations}
\tablewidth{0pt}
\tablehead{
\colhead{Field}
&\colhead{1st Obs.}
&\colhead{2nd Obs.}
&\colhead{3rd Obs.}
&\colhead{4th Obs.}
}
\startdata
M31\_1 & 2012 Oct 9 & 2012 Oct 11 & 2012 Nov 7 & 2012 Dec 7 \\
M31\_2 & 2012 Nov 7 & 2012 Nov 8 & 2012 Dec 6 & 2012 Dec 11 \\
\hline
M33\_1 & 2012 Oct 10 & 2012 Nov 6 & 2012 Nov 7 & 2012 Dec 8 \\
M33\_2 & 2012 Nov 8 & 2012 Dec 8 & 2012 Dec 10 & \nodata \\
\enddata
\end{deluxetable}

\begin{deluxetable}{l l l c c c c c  c c c c c}
\rotate
\tablecaption{\label{tab:candidates} Newly Found M33 WR and Of-type Stars}
\tablewidth{0pt}
\tabletypesize{\footnotesize}
\tablehead{
&
&
&\multicolumn{2}{c}{He II $\lambda 4686$}
&
&\multicolumn{2}{c}{C IV $\lambda 5606$} \\ \cline{4-5} \cline{7-8}
\colhead{Star}
&\colhead{$\rho$\tablenotemark{a} }
&\colhead{Type}
&\colhead{$\log$(-EW)}
&\colhead{FWHM(\AA)}
&
&\colhead{$\log$(-EW)}
&\colhead{FWHM(\AA)}
&\colhead{$V$\tablenotemark{b}}
&\colhead{$m_{\lambda 4750}$\tablenotemark{c}}
&\colhead{$M_V$\tablenotemark{d}}
&\colhead{$M_{\lambda 4750}$\tablenotemark{d}}
&\colhead{OB\tablenotemark{e}}
}
\startdata
J013302.73+301131.6 &     0.97  &   WN2.5+neb &1.9& 27&& \nodata & \nodata & 20.00 & 19.65& $-5.0$& $-5.4$& Fld\\
J013404.07+304658.3 &     0.26  &   WN6       &1.0&     14 && \nodata & \nodata & 19.52 & 19.40& $-5.5$&$-5.7$& (71)\\
J013411.45+303637.3 &     0.31  &   WN4       &1.8&     26&&  0.7& 36& 21.34 & 21.24& $-3.6$ & $-3.8$ & (100)\\             
J013432.60+304211.3 &     0.46  &   O6.5 II(f)&\nodata&\nodata&&\nodata&\nodata& 19.34 & 19.10& $-5.6$& $-6.0$ & (91)\\
J013434.26+304637.8 &     0.47  &   WN6       &1.4&16 &&\nodata  & \nodata & 20.48 & 20.71 & $-4.5$& $-4.3$& 84\\
J013442.41+305019.0 &     0.58  &   WN4.5     &1.5& 17&&  0.5&  27& 20.95 & 21.19 & $-4.0$ & $-3.9$& Fld\\
\enddata
\tablenotetext{a}{Distance from the center within the plane of M33, normalized to the $D_25$ isophotoal radius of 30\farcm8 assuming $\alpha_{2000}=01^h33^m50\fs89$, $\delta_{2000}=30^\circ 39\arcmin36\farcs8$, an inclination of 56$^\circ$ and a position angle of the major axis of 23$^\circ$, following Kwitter \& Aller 1981 and Zaritsky et al.\ 1989.}
\tablenotetext{b}{From Massey et al.\ 2006.}
\tablenotetext{c}{AB magnitude through CT filter centered at 4750 \AA\ and calibrated using the values from Massey \& Johnson 1998.}
\tablenotetext{d}{Absolute magnitudes computed assuming a true distance modulus of 24.60 (830 Mpc) and adopting an average reddening of $E(B-V)=0.12$ based on Massey et al.\ 2007.  Adopting $R_V=3.1$ leads to an $A_V=0.37$ mag and $A_{m_{4750}}=0.45$ mag.}
\tablenotetext{e}{OB association as defined by Humphreys \& Sandage 1980.  Parenthesis means the star is just outside the boundaries of the association, while ``Fld" implies it is a field star, not in a cataloged OB association.}
\end{deluxetable}

\begin{deluxetable}{l l l l l}
\tablecaption{\label{tab:Sigma}$E/I$ Values\tablenotemark{*}}
\tablewidth{0pt}
\tablehead{
\colhead{Star Name}
&\colhead{$\lambda$ (\AA)}
&\colhead{$E/I_{\rm{fxcor}}$}
&\colhead{$E/I_{\rm{hand}}$}
&\colhead{$N$}
}
\startdata
J004410.91+411623 & 4686 & 12.2 & 10.5 & 5\\
J004234.42+413024 & 4686 & 12.0 & 10.8 & 5\\
J013505.37+304114 & 4650 & 10.8 &  9.2 & 3\\
J004147.24+410647 & 4686 &  9.4 &  9.3  & 6\\
J004425.83+415019 & 4686 &  9.0 &  9.5  & 5\\
J013342.53+303314 & 4650 &  7.9 & 10.5 & 3\\
J004517.89+415209 & 4686 &  6.1 &  5.1 & 5\\
J004506.50+413425 & 4686 &  6.0 &  5.5  & 5\\
J004031.67+403909 & 4686 &  5.9 &  4.0  & 5\\
J013402.93+305126 & 4686 &  5.7 &  6.0 & 9 \\
J013312.61+304531 & 4686 &  5.5 &  4.5  & 6\\
J004537.10+414201 & 4650 &  5.5 &  4.3 & 5 \\
J004141.81+403711 & 4686 &  5.4 &  4.5  & 3\\
J013359.60+303435 & 4686 &  4.9 &  4.6 & 5 \\
J004005.65+405848 & 4686 &  4.8 &  3.7  & 6\\
J004032.58+403550 & 4686 &  4.8 &  3.5 & 5 \\
J004148.27+411739 & 4686 &  4.5 &  2.4  & 9\\
J013305.67+302857 & 4686 &  4.4 &  4.3 & 4 \\
J004347.01+411238 & 4686 &  4.4 &  3.3  & 5\\
J004026.23+404459 & 4686 &  4.4 &  3.1  & 6\\
J013334.28+303347 & 4686 &  4.2 &  3.9  & 3\\
J013307.68+303315 & 4686 &  4.1 &  3.9  & 8\\
J013327.76+303150 & 4686 &  4.1 &  3.5  & 3\\
J013303.71+302326 & 4686 &  3.9 &  4.5  & 7\\
X004256.05+413543 & 4650 &  3.9 &  3.2 & 6\\
J013417.21+303334 & 4686 &  3.9 &  2.9 & 5 \\
J003939.97+403450 & 4650 &  3.8 &  3.8  & 6\\
J013318.50+302658 & 4650 &  3.8 &  3.3  & 3\\
J013423.02+304650 & 4686 &  3.8 &  2.4  & 9\\
J004331.17+411203 & 4686 &  3.6 &  2.6  & 5\\
J013340.19+303134 & 4650 &  3.5 &  3.3  & 4\\
J013416.28+303646 & 4650 &  3.5 &  3.1  & 4\\
J013311.29+303146 & 4686 &  3.4 &  3.4 & 8 \\
J004517.56+413922 & 4686 &  3.4 &  2.4 & 5 \\
J013357.20+303512 & 4686 &  3.4 &  2.4  & 3\\
J004203.94+412554 & 4650 &  3.3 &  2.9 & 5 \\
J004043.28+403525 & 4650 &  3.3 &  2.5  & 5\\
J013415.85+305522 & 4686 &  3.3 &  2.2  & 7\\
J013444.28+303757 & 4686 &  3.3 &  2.1  & 8\\
J004102.04+410446 & 4686 &  3.2 &  4.2 & 5 \\
J003948.84+405256 & 4686 &  3.2 &  2.7  & 5\\
J013335.23+310037 & 4686 &  3.2 &  2.6  & 4\\
J004444.87+412800 & 4686 &  3.2 &  2.2 & 5 \\
J013240.82+302454 & 4686 &  3.1 &  3.5  & 7\\
J013401.30+304004 & 4650 &  3.1 &  2.6  & 3\\
J004114.95+404448 & 4686 &  3.1 &  2.4  & 5\\
J013326.67+304040 & 4650 &  3.1 &  2.3  & 7\\
J004520.80+415100 & 4686 &  3.1 &  2.3  & 5\\
J013406.80+304727 & 4686 &  3.1 &  2.1 & 5 \\
J013426.96+305256 & 4686 &  3.1 &  1.9 & 4 \\
J013407.85+304145 & 4686 &  3.0 &  3.4 & 7 \\
J004436.22+412257 & 4686 &  3.0 &  3.1  & 5\\
J004455.82+412919 & 4686 &  3.0 &  3.0 & 5 \\
J004247.12+405657 & 4650 &  3.0 &  2.3  & 5\\
J004024.33+405016 & 4686 &  3.0 &  2.2  & 5\\
J004420.58+415412 & 4686 &  3.0 &  2.1 & 5 \\
J013340.07+304238 & 4650 &  2.9 &  2.0  & 3\\
J013340.28+304053 & 4650 &  2.8 &  1.8  & 4\\
J013257.32+304418 & 4686 &  2.8 &  1.5  & 8\\
J013314.34+302955 & 4686 &  2.7 &  3.0& 3 \\
J013302.28+301119 & 4686 &  2.7 &  2.4 & 4 \\
J013350.71+305636 & 4686 &  2.7 &  1.8 & 5 \\
J013345.58+303451 & 4686 &  2.7 &  1.6& 3 \\
J013353.80+303528 & 4686 &  2.7 &  1.6 & 3 \\
J013336.67+304302 & 4686 &  2.7 &  1.4 & 5\\
J004511.27+413815 & 4650 &  2.6 &  2.5 & 5\\
J013232.07+303522 & 4686 &  2.6 &  2.2 & 5\\
J013308.56+302805 & 4686 &  2.6 &  1.9& 4 \\
J004349.72+411243 & 4686 &  2.6 &  1.4& 5 \\
J013241.95+304024 & 4650 &  2.5 &  3.2 & 7\\
J004425.10+412050 & 4650 &  2.5 &  1.9& 5 \\
J004306.11+413813 & 4686 &  2.5 &  1.5 & 5\\
J013440.42+304321 & 4686 &  2.5 &  1.4 & 8\\
J013255.91+303155 & 4650 &  2.4 &  2.7& 8 \\
J013340.32+304600 & 4686 &  2.4 &  2.6 & 5\\
J004344.48+411142 & 4686 &  2.4 &  2.1 & 6 \\
J004109.46+404907 & 4686 &  2.4 &  2.0& 5 \\
J004256.85+413837 & 4686 &  2.4 &  1.5 & 5\\
J013256.84+302724 & 4686 &  2.3 &  2.0 & 4\\
J013332.82+304146 & 4686 &  2.3 &  1.4& 3 \\
J004336.51+412315 & 4686 &  2.3 &  1.2& 5 \\
J004304.34+412223 & 4686 &  2.3 &  1.1& 5 \\
J013307.50+304258 & 4686 &  2.2 &  3.2 & 8\\
J004145.33+412104 & 4686 &  2.2 &  2.4 & 9\\
J004023.02+404454 & 4686 &  2.2 &  2.2 & 5\\
J004406.68+411612 & 4686 &  2.2 &  1.8 & 5\\
J013438.98+304119 & 4650 &  2.2 &  1.7 & 4\\
J013337.81+302831 & 4686 &  2.2 &  1.1 & 7\\
J013355.94+302732 & 4686 &  2.2 &  0.8 & 8\\
J004426.32+413419 & 4686 &  2.1 &  2.6 & 6\\
J004036.76+410104 & 4686 &  2.1 &  1.8& 5 \\
J013334.93+310042 & 4686 &  2.1 &  1.7& 5 \\
J013418.37+303837 & 4686 &  2.1 &  1.5& 4 \\
J013443.51+304919 & 4686 &  2.1 &  1.4& 5 \\
J013442.41+305019 & 4686 &  2.1 &  1.4 & 6\\
J004353.34+414638 & 4686 &  2.1 &  1.2& 5 \\
J013256.35+303535 & 4686 &  2.1 &  1.0& 5 \\
J003935.63+402811 & 4586 &  2.0 &  1.3 & 5\\
J013355.94+303407 & 4650 &  2.0 &  1.3 & 3\\
J013312.44+303848 & 4686 &  2.0 &  1.2& 4 \\
J004143.00+411859 & 4686 &  2.0 &  1.1 & 9\\
J013359.79+305150 & 4686 &  2.0 &  0.8& 8 \\
J013312.54+303900 & 5801 &  1.9 &  2.9 & 4\\
J013245.84+302019 & 4650 &  1.9 &  2.4 & 8\\
J004020.44+404807 & 4650 &  1.9 &  2.1 & 5\\
J013355.33+302001 & 4686 &  1.9 &  2.0 & 8\\
J013510.27+304522 & 4686 &  1.9 &  1.9 & 4\\
J013338.20+303112 & 4650 &  1.9 &  1.6 & 8\\
J013241.39+303416 & 4686 &  1.9 &  1.5 & 8\\
J013400.57+303809 & 4686 &  1.9 &  1.4 & 7\\
J004443.10+412619 & 4686 &  1.9 &  1.3& 5 \\
J013300.20+303015 & 4686 &  1.9 &  1.2& 3 \\
J013346.55+303700 & 4650 &  1.9 &  1.1 & 4\\
J013245.74+303854 & 4686 &  1.9 &  1.1 & 4\\
J013427.30+305229 & 4686 &  1.9 &  1.1 & 3\\
J013332.64+304127 & 4686 &  1.9 &  0.7& 4 \\
J013311.44+304856 & 4686 &  1.9 &  0.6& 7 \\
J013348.85+303949 & 4686 &  1.9 &  0.5& 3 \\
J004444.00+412739 & 4650 &  1.8 &  1.6 & 5\\
J004509.18+414021 & 4650 &  1.8 &  1.4& 5 \\
J004539.15+414754 & 4686 &  1.8 &  1.0& 5\\
J004321.48+414155 & 4686 &  1.8 &  1.0 & 6\\
J004445.90+415803 & 4686 &  1.8 &  0.6 & 5\\
J013434.26+304637 & 4686 &  1.8 &  0.5 & 3\\
J013316.17+304751 & 4650 &  1.7 &  1.7 & 5\\
J013401.73+303620 & 4650 &  1.7 &  1.5 & 3\\
J004238.90+410002 & 4686 &  1.7 &  1.3 & 6\\
J004451.98+412911 & 4686 &  1.7 &  0.9 & 5\\
J004514.10+413735 & 4686 &  1.6 &  2.1 & 5\\
J013411.45+303637 & 4686 &  1.6 &  1.4 & 3\\
J013324.04+305030 & 4686 &  1.6 &  1.3& 9 \\
J013356.33+303420 & 4686 &  1.6 &  0.8 & 5\\
J004143.09+404045 & 4686 &  1.6 &  0.8 & 5\\
J004434.57+412424 & 4686 &  1.6 &  0.7 & 5\\
J013347.67+304351 & 4686 &  1.6 &  0.6 & 7\\
J013355.87+304528 & 4686 &  1.6 &  0.4 & 5\\
J013346.20+303436 & 4686 &  1.5 &  1.8& 4 \\
J004021.13+403520 & 4686 &  1.5 &  1.5 & 5\\
J013447.32+310748 & 4686 &  1.5 &  1.4& 5 \\
J013409.12+303907 & 4650 &  1.5 &  1.4 & 4\\
J013421.97+303314 & 4686 &  1.5 &  1.3 & 5\\
J004403.39+411518 & 4686 &  1.5 &  1.2 & 5\\
J004502.78+415533 & 4686 &  1.5 &  1.0 & 5\\
J013337.34+303527 & 4686 &  1.5 &  0.8 & 5\\
J013330.36+303128 & 4686 &  1.5 &  0.7 & 5\\
J013425.11+301950 & 4686 &  1.5 &  0.6 & 5\\
J013507.24+304500 & 4650 &  1.4 &  1.3 & 5\\
J013352.75+304444 & 4686 &  1.4 &  1.2 & 3\\
J004453.52+415354 & 4686 &  1.4 &  1.1 & 5\\
J013316.48+303221 & 4686 &  1.4 &  1.1 & 3\\
J004126.11+411220 & 4686 &  1.4 &  1.0 & 5 \\
J013416.35+303712 & 4686 &  1.4 &  0.6& 3 \\
J004437.61+415203 & 4686 &  1.4 &  0.6 & 5\\
J004130.37+410500 & 4686 &  1.4 &  0.5 & 5\\
J004510.39+413646 & 4686 &  1.4 &  0.3 & 5\\
J004056.49+410308 & 4686 &  1.4 &  0.2& 5 \\
J013419.16+303127 & 4686 &  1.3 &  2.0 & 8\\
J004415.77+411952 & 4650 &  1.3 &  1.6 & 5\\
J004107.31+410417 & 4650 &  1.3 &  1.5 & 5\\
J004002.84+403852 & 4686 &  1.3 &  1.5 & 5\\
J004436.52+412202 & 4686 &  1.3 &  1.2 & 5\\
J013358.69+303526 & 4686 &  1.3 &  0.9& 3 \\
J013408.23+305234 & 4686 &  1.3 &  0.8 & 5\\
J013327.26+303909 & 4686 &  1.3 &  0.7& 3 \\
J013314.56+305319 & 4686 &  1.3 &  0.7 & 8\\
J013312.95+304459 & 4686 &  1.3 &  0.6 & 7\\
J013352.01+304023 & 4650 &  1.2 &  1.7 & 3\\
J004530.61+414639 & 4650 &  1.2 &  1.5 & 5\\
J013438.18+304953 & 4686 &  1.2 &  1.5 & 4\\
J013444.61+304445 & 4650 &  1.2 &  1.4 & 8\\
J013350.23+303342 & 4686 &  1.2 &  1.4 & 4\\
J004115.95+410906 & 4686 &  1.2 &  1.3 & 5\\
J013312.21+302740 & 4686 &  1.2 &  1.2& 3 \\
J004435.15+412545 & 4650 &  1.2 &  1.2 & 5\\
J003911.04+403817 & 4650 &  1.2 &  1.2 & 5\\
J004412.44+412941 & 4650 &  1.2 &  1.1 & 5\\
J013354.40+303453 & 4650 &  1.2 &  0.8 & 4\\
J013339.30+303554 & 4686 &  1.2 &  0.8 & 3\\
J013340.21+303551 & 4650 &  1.2 &  0.7 & 4\\
J004324.95+411803 & 4686 &  1.2 &  0.5 & 5\\
J013352.71+304502 & 4650 &  1.1 &  1.8 & 4\\
J013341.65+303855 & 4686 &  1.1 &  1.7 & 2\\
J013233.24+302652 & 4686 &  1.1 &  1.4 & 7\\
J013410.72+305240 & 4686 &  1.1 &  1.2 & 5\\
J004455.63+413105 & 4650 &  1.1 &  1.1 & 5\\
J013303.21+303408 & 4686 &  1.1 &  1.0 & 8\\
J013326.60+303550 & 4686 &  1.1 &  0.9& 8 \\
J013352.83+304347 & 4686 &  1.1 &  0.6 & 2\\
J004430.04+415237 & 4686 &  1.1 &  0.0& 5 \\
J004327.92+414207 & 4686 &  1.0 &  1.7 & 5\\
J004019.47+405224 & 4650 &  1.0 &  1.5 & 6\\
J004211.16+405648 & 4686 &  1.0 &  1.0& 5 \\
J013421.21+303758 & 4686 &  1.0 &  1.0 & 5\\
J013351.84+303328 & 4650 &  1.0 &  1.0 & 4\\
J013431.45+305716 & 4686 &  1.0 &  0.9 & 5\\
J013310.77+302734 & 4686 &  1.0 &  0.8 & 5\\
J004240.81+410241 & 4650 &  1.0 &  0.5 & 5\\
J013419.58+303801 & 4686 &  1.0 &  0.5 & 4\\
J013350.26+304134 & 4686 &  1.0 &  0.5 & 4\\
J013343.19+303906 & 4686 &  1.0 &  0.4 & 4\\
J013307.80+302951 & 4686 &  0.9 &  1.9 & 3\\
J004134.99+410552 & 4650 &  0.9 &  1.6 & 5\\
J013339.70+302101 & 4686 &  0.9 &  1.3 & 7\\
J004029.27+403916 & 4650 &  0.9 &  1.3 & 5\\
J013404.07+304658 & 4686 &  0.9 &  1.2& 7 \\
J004144.47+404517 & 4650 &  0.9 &  1.2 & 5\\
J004537.05+414302 & 4686 &  0.9 &  1.1 & 5\\
J004214.36+412542 & 4650 &  0.9 &  0.9 & 5\\
J013433.22+310019 & 4686 &  0.9 &  0.7& 8 \\
J004406.39+411921 & 4650 &  0.9 &  0.7 & 4\\
J004422.24+411858 & 4686 &  0.9 &  0.6 & 5\\
J004242.03+412314 & 4650 &  0.9 &  0.4 & 5\\
J013432.78+304703 & 4650 &  0.8 &  1.8 & 3\\
J013345.99+303602 & 4686 &  0.8 &  1.8 & 3\\
J004408.58+412121 & 4650 &  0.8 &  1.2 & 5\\
J013344.40+303845 & 4650 &  0.8 &  0.9 & 4\\
J004341.72+412304 & 4650 &  0.8 &  0.8 & 5\\
J013347.15+303702 & 4650 &  0.8 &  0.8 & 3\\
J004257.62+413727 & 4686 &  0.8 &  0.7 & 5\\
J013432.18+304903 & 4686 &  0.8 &  0.6 & 5\\
J004330.76+412734 & 4686 &  0.8 &  0.6 & 5\\
J004334.92+410953 & 4686 &  0.8 &  0.5 & 5\\
J013400.90+303918 & 4686 &  0.8 &  0.4 & 4\\
J004154.62+404713 & 4650 &  0.7 &  1.7 & 5\\
J013353.25+304413 & 4650 &  0.7 &  1.7 & 4\\
J013429.56+304145 & 4650 &  0.7 &  1.3 & 5\\
J013302.67+303120 & 4686 &  0.7 &  1.3 & 5\\
J004410.17+413253 & 4650 &  0.7 &  1.0 & 5\\
J013347.83+303338 & 4686 &  0.7 &  1.0 & 4\\
J013232.13+303514 & 4686 &  0.7 &  1.0 & 4\\
J004059.13+403652 & 4650 &  0.7 &  0.8 & 6\\
J013347.96+304506 & 4686 &  0.7 &  0.6 & 4\\
J013343.33+304450 & 4686 &  0.7 &  0.4 & 5\\
J004413.06+411920 & 4686 &  0.7 &  0.3 & 5\\
J004034.17+404340 & 4686 &  0.7 &  0.3 & 5\\
J004213.21+405051 & 4650 &  0.6 &  1.1 & 6\\
J004524.26+415352 & 4650 &  0.6 &  0.9 & 5\\
J013408.90+304732 & 4686 &  0.6 &  0.8 & 6\\
J013315.55+304514 & 4686 &  0.6 &  0.8 & 5\\
J013335.47+304220 & 4650 &  0.5 &  1.7 & 7\\
J013433.82+304656 & 4650 &  0.5 &  0.9 & 5\\
J013257.88+303549 & 4650 &  0.5 &  0.6 & 4\\
J004249.84+410215 & 4650 &  0.5 &  0.2 & 5\\
J004449.41+413020 & 4686 &  0.4 &  1.2 & 5\\
J013341.83+304154 & 4650 &  0.4 &  0.6 & 3\\
J013419.68+303343 & 4686 &  0.3 &  1.4 & 4\\
J013304.98+303159 & 4686 &  0.3 &  0.4 & 3\\
J013340.23+304102 & 4686 &  0.2 &  2.0 & 3\\
J013352.71+303907 & 4686 &  0.2 &  1.2 & 3\\
J013411.14+304637 & 4650 &  0.2 &  0.8 & 8\\
\enddata
%\tablenotetext{*}{The full version of this table can be found online.}
\end{deluxetable}

\begin{deluxetable}{l c c c c c c c c c c c c}
\tabletypesize{\footnotesize}
\tablecaption{\label{tab:BinFreq} Binary Frequency\tablenotemark{a}}
\tablewidth{0pt}
\tablehead{
\colhead{Region}
&\colhead{$\bar{\rho}$}
&\multicolumn{2}{c}{log$\frac{O}{H}+12$}
&\colhead{Total}
&\multicolumn{3}{c}{Total \% Binary}
&
&\multicolumn{3}{c}{\% Binary $\frac{E}{I} > 2$} \\ \cline{3-4} \cline{6-8} \cline{10-11}
&
&\colhead{Value}
&\colhead{Ref.\tablenotemark{b}}
&\colhead{\# Stars}
&\colhead{$\frac{E}{I} > 2$}
&\colhead{$\frac{E}{I} > 3$}
&\colhead{$\frac{E}{I} > 4$}
&
&\colhead{WNs}
&\colhead{WCs}
}
\startdata
M31 (all)                           & 0.53 & 8.9 & 1 & 106 &$44\pm8$ &$28\pm6$ &$13\pm4$ && $57\pm12$ & $27\pm9$ \\
M33 ($\rho<0.25$)             & 0.16 & 8.7 & 2 & 44 & $23\pm8$ & $11\pm5$ & $5\pm3$ && $20\pm10$ & $26\pm13$\\
M33 ($0.25\le \rho <0.50$)& 0.38  & 8.4 & 2 & 46 & $46\pm12$ & $26\pm8$ & $9\pm5$ && $47\pm14$ & $40\pm24$\\
M33 ($\rho\ge 0.5$)           & 0.69  & 8.3 & 2 & 54 & $44\pm11$ & $17\pm6$ & $6\pm3$ && $43\pm12$ & $50\pm27$\\
\enddata
\tablenotetext{a}{Errors on the percentage of binaries are statistical and assume that the uncertainty on the number $N$ is simply $\sqrt{N}$.}
\tablenotetext{b}{References for oxygen abundances: 1--Sanders et al.\ 2012. 2--Magrini et al.\ 2007.}
\end{deluxetable}

\end{document}